\documentclass[conference]{IEEEtran}

\usepackage{txfonts}
\usepackage[T1]{fontenc}
\usepackage{times}
\usepackage{helvet}
\usepackage{courier}
\usepackage{url}
\usepackage{ifpdf}
\usepackage{multirow}
\usepackage{array}
\usepackage{rotating}
\usepackage{color}
\usepackage{cite}
\usepackage{graphicx}
\usepackage{graphics}
\usepackage[table]{xcolor}
\definecolor{lightgray}{gray}{0.9}

\begin{document}
%
\title{Sensitivity and Reliability in Incomplete Networks: Centrality Metrics to Community Scoring Functions}

\author{\IEEEauthorblockN{Soumya Sarkar}
\IEEEauthorblockA{Dept. of Comp. Sc. \& Engg.\\
IIT Kharagpur\\
West Bengal, India -- 721302\\
Email: soumya015@iitkgp.ac.in}\\
\IEEEauthorblockN{Sanjukta Bhowmick}
\IEEEauthorblockA{Dept. of Comp. Sc.\\
Univ. of Nebraska, Omaha\\
Omaha, NE 68106, USA\\
Email: sbhowmick@unomaha.edu}
\and
\IEEEauthorblockN{Suhansanu Kumar}
\IEEEauthorblockA{Dept. of Comp. Sc.\\
UIUC, Champaign, IL 61820, USA\\
Email: suhansanu1@gmail.com}\\
\IEEEauthorblockN{Animesh Mukherjee}
\IEEEauthorblockA{Dept. of Comp. Sc. \& Engg.\\
IIT Kharagpur\\
West Bengal, India -- 721302\\
Email: animeshm@cse.iitkgp.ernet.in}
}


%


\maketitle

\begin{abstract}
In this paper we evaluate the effect of noise on community scoring and centrality-based parameters with respect to two different aspects of network analysis: (i) sensitivity, that is how the parameter value changes as edges are removed and (ii) reliability in the context of message spreading, that is how the time taken to broadcast a message changes as edges are removed. 

Our experiments on synthetic and real-world networks and three different noise models demonstrate that for both the aspects over all networks and all noise models, {\em permanence} qualifies as the most effective metric. For the sensitivity experiments closeness centrality is a close second. For the message spreading experiments, closeness and betweenness centrality based initiator selection closely competes with permanence. This is because permanence has a dual characteristic where the cumulative permanence over all vertices is sensitive to noise but the ids of the top-rank vertices, which are used to find seeds during message spreading remain relatively stable under noise. 
\end{abstract}

\section{Introduction}

Network analysis has become an ubiquitous tool for understanding the behavior of various complex systems~\cite{Mitchell06complexsystems:}. The vertices in the network represent the entities of the complex system and the edges represent their pairwise interactions.

However, in the practical context, due to the limitations in data gathering, not all interactions can be observed. Consequently, the network can be potentially incomplete, as manifested by missing edges. It is therefore important to determine the effect of this incompleteness or noise on different network parameters and rank them according to how they behave under noise. 

In this paper we study the effect of noise on two important classes of  network analysis metrics -- (i) centrality measures and (ii) community scoring functions.  Centrality measures are key to applications that rely on node ranking, and the community scoring functions determine the quality of clusters/communities  that are used in many applications requiring unsupervised classification. 

We evaluate these metrics based on two orthogonal qualities. The first is {\bf sensitivity}, that is whether the change in the value of the metric is commensurate with percentage of edges removed. If a metric is sensitive then it can serve as a good indicator of how much the network has been changed (see Section ~\ref{Section3}). 

The second metric is {\bf reliability}, that is whether certain operations in the network can be performed efficiently in spite of the missing edges. If a metric is reliable then it guarantees good performance even under noise. Here we select message spreading as the candidate operation. The seed nodes that initiate message spreading are selected from the high valued entities of different metrics. A metric has high reliability if the time for message spreading under noise does not significantly degrade if its high valued entities are selected as seeds (see Section~\ref{Section4}). In contrast to previous work~\cite{borgatti2006robustness,yan2011finding,kossinets2006effects}, which focused on single noise models and primarily on centrality metrics, to the best of our knowledge this is the first comparative study encompassing several centrality and community-scoring parameters and different types of noise models.


{\bf Overview of Experiments (Section~\ref{Section2})} Among the centrality measures we consider  closeness, betweenness and Pagerank and among the community scoring functions we consider modularity, cut-ratio and conductance.  We also include a third type of metric, {\em permanence}~\cite{chakraborty2014permanence}. Although permanence is a community scoring metric, unlike the others it is vertex based. Therefore permanence can also be considered as a centrality measure. 

We apply three different noise models on real-world and synthetic networks. We empirically evaluate the above metrics to estimate their sensitivity to varying levels of noise. We also measure their reliability by observing whether high valued vertices of these metrics can serve as effective seeds for message spreading.

In all our experiments, we  ensure that in spite of the noise, the underlying community structure is not significantly disrupted from its original form and the giant component of the network remains connected. Nevertheless, as we shall see, even this constrained noise can significantly affect the analysis.

\noindent{\bf Key Results}
(i) For both the objectives -- sensitivity and reliability and for all the given noise models and networks, permanence proves to be the most sensitive and most reliable metric in majority of the cases. (ii) The other centrality metrics can be ranked in a partial order. The only other metric that exhibits sensitivity is closeness. For reliability, when a difference in the performance can be observed, closeness and betweenness also show high reliability. (iii) For all sensitivity experiments, and for most reliability experiments, the partial ordering of metrics is relatively independent of the noise model and type of network. Community scoring metrics,apart from permanence, are not sensitive. 

{\bf Rationale for the behavior of permanence (Section ~\ref{Section5}).} At a quick glance it would seem that sensitivity and reliability are mutually opposing properties. Sensitivity is used as an indicator of noise, whereas reliability is used to guarantee good performance in spite of noise. It is therefore surprising that permanence is both the most sensitive as well as the most reliable among all the metrics that we investigate.  

We believe that this is because permanence encompasses both community-like and centrality-like properties. While the cumulative value of permanence is sensitive to the level of noise, satisfying the sensitivity criterion, its high rank vertices are stable under noise and therefore serve as effective seeds in the noisy versions of the networks. We compute the Jaccard Index (JI) for the high ranked vertices between the original and noisy networks. Permanence exhibits the highest JI and therefore the set of its high ranked vertices change the least.

\begin{figure}
\begin{tabular}{l}
\includegraphics[scale=.2]{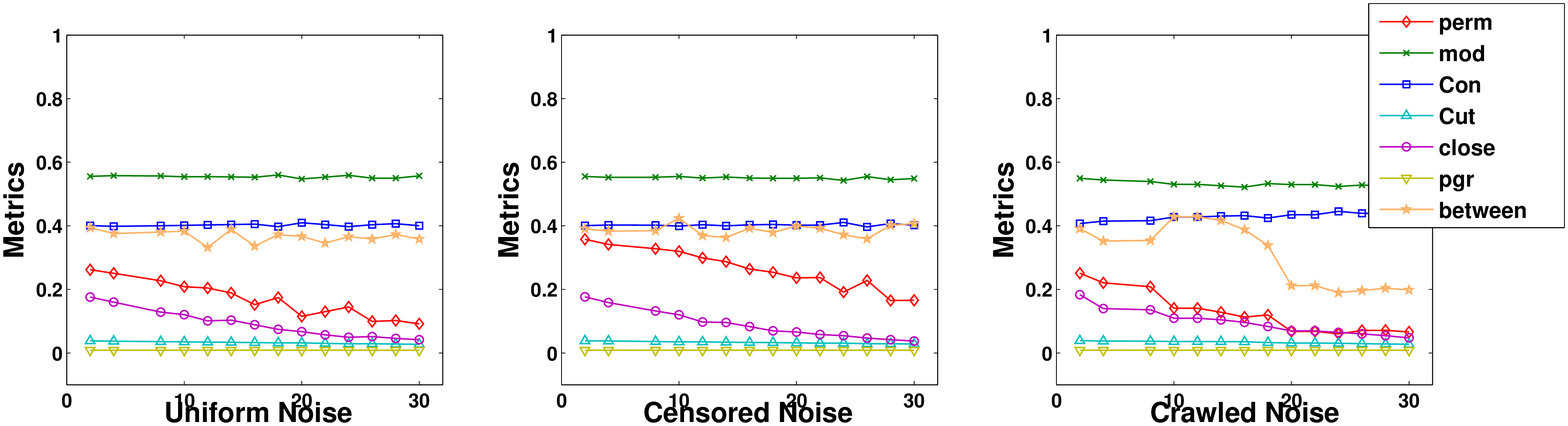} \\ \includegraphics[scale=.2]{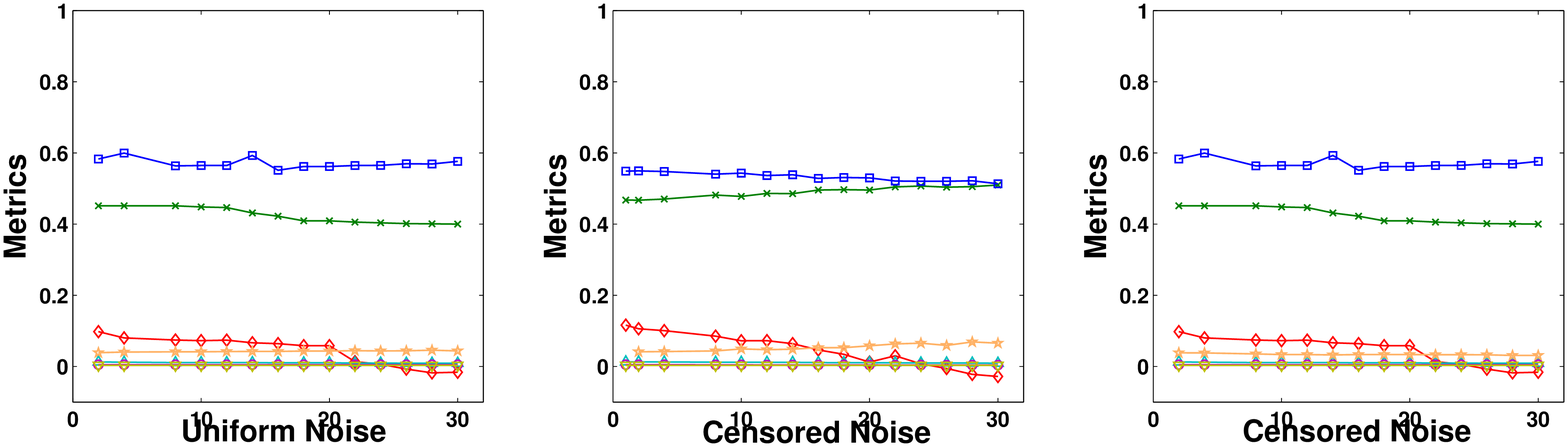}
\\
\includegraphics[scale=.2]{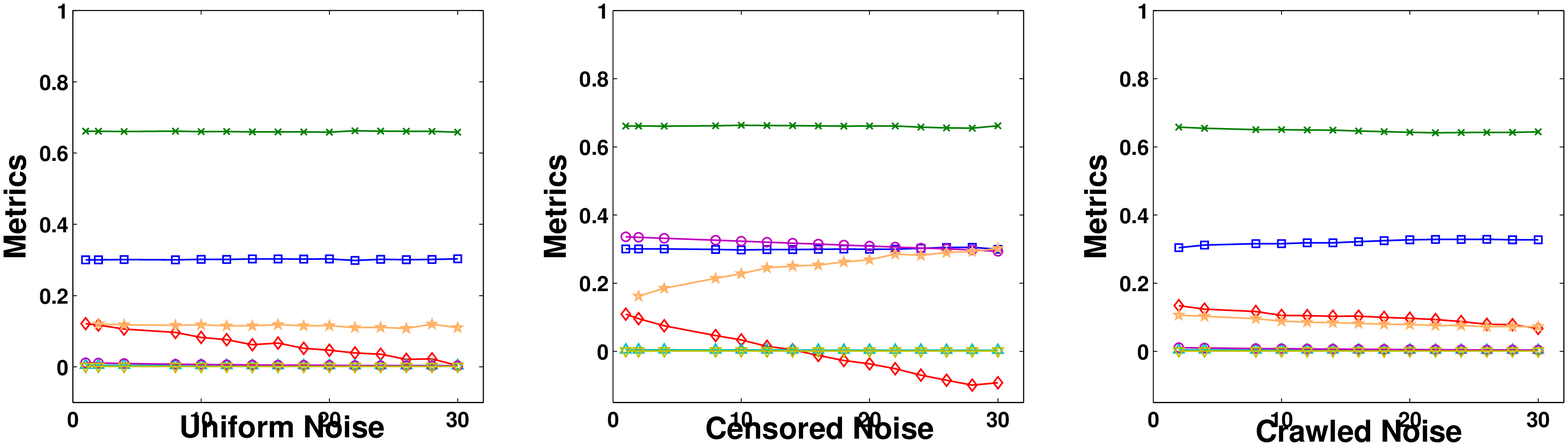}
\end{tabular}
  \caption{Sensitivity of the different quality metrics for varying levels of noise (in steps of 2\%) with x-axis as the noise level and y-axis as the metric values. The first and the second panels are for the football and railway the networks respectively. The third panel is for the LFR network ($\mu=0.3$).}  
  \label{sense}
\vspace{-.5cm}
\end{figure}


\section{Experimental setup}\label{Section2}

\noindent {\bf Datasets.}
Here is a  brief description of the different networks we used (see Table~\ref{tab:my_label} for properties of real-world networks). 

\noindent{\bf LFR Benchmark:} We use the benchmark {\bf LFR model} \cite{lancichinetti2009benchmarks} that generates different networks and ground-truth communities of various quality. We use $n=1000$, $\mu=0.3$ keeping all other parameters to the default values of the original implementation~\footnote{\url{https://sites.google.com/site/santofortunato/inthepress2}}.


\noindent{\bf Railway:} has been taken from Ghosh et al.~\cite{ghosh2011statistical}. 

\noindent{\bf Football:} has been taken from Girvan et al.~\cite{girvan2002community}. 

 \begin{table}
\centering
\resizebox{6.5cm}{!}{
\begin{tabular}{|c | c | c | c | c | c | c | c |} 
\hline
Network & \#Nodes & \#Edges & \textless k\textgreater & $k_{max}$ & $|c|$ & $n_{c}^{max}$ & $n_{c}^{min}$   \\ \hline
Football & 115 & 613 & 10.57 & 12 & 12 & 5 & 13 \\ \hline
Railway  & 301 & 1224 & 6.36  & 48  & 21 & 1 & 46  \\ \hline

\end{tabular}}
\caption{Dataset statistics. $|c|$ denotes the number of communities in the ground-truth, $n_{c}^{min}$ and $n_{c}^{max}$ denote the number of nodes in the smallest and the largest size communities respectively.}
\label{tab:my_label}
\vspace{-.8cm}
\end{table}


\noindent{\bf Noise Models.}
We experiment with three noise models -- \textbf{uniform}, \textbf{crawled} and \textbf{censored} (see~\cite{yan2011finding} for detailed description), -- to simulate  real-world sources of noise. We do not allow formation of disconnected components while introducing noise. We vary noise levels in steps of 2\% from $0$ to $30$ in all our experiments.  

\noindent{\bf Metrics.} Our set of network parameters for evaluation include community-scoring metrics namely,  modularity, cut-ratio, and conductance, and centrality metrics namely betweenness, closeness and  Pagerank. For the definitions of these metrics the reader is referred to~\cite{newman2010networks}. We also include a recently introduced metric permanence~\cite{chakraborty2014permanence}, that serves both as a community scoring function as well as a measure of centrality. 

\section{Sensitivity of the metrics}\label{Section3}

A {\em sensitive} parameter is one whose change is commensurate with the amount of noise applied. For small amounts of noise, the change in the parameter values should be low, whereas, as the noise increases, the change should be much higher. A sensitive parameter can function as a good indicator of whether a network significantly changed from its original topology.\\
{\bf Our goal is to rank the network parameters by the extent to which they are sensitive  to the noise level.} 

{\bf Methodology.}  We apply the three noise models on the one synthetic LFR network ($\mu = 0.3$), and two real-world, railway and football, networks. For each increasing level of noise we compute the value of the parameters. For the vertex-based metrics we take the average over all vertices. We compute the value of the community-scoring parameters based on the ground-truth community assignment from the original network. Our rationale is that because community detection is expensive, therefore, re-computing the community after each noise addition would defeat the purpose of quickly ascertaining the change in the network. Further, our selected noise level is low enough such that it does not significantly change the original ground-truth community. Our results are averaged over ten simulation runs. 



 {\bf Results.} The results in Fig.~\ref{sense} show that the change in  permanance has the highest slope with respect to increasing noise. This indicates that permanence is most sensitive to noise as compared to the other parameters.  However, there are some cases, e.g., the football network where the closeness centrality is also quite sensitive. While betweeness is slightly sensitive in the LFR networks, it shows an opposite trend, i.e. increase in value with noise for censored noise in the real-world networks.  The rest of the metrics remain constant. We report the range of the average of each metric obtained for each noise model as a tuple -- (average metric value at 2\% noise level, average metric value at 30\% noise level) -- in Table~\ref{range}. In this table, permanence shows the largest separation.

\begin{table}
\centering
\resizebox{8.5cm}{!}{
\begin{tabular}{l}
\begin{tabular}{llll}
\hline
Metrics       & Railway           & Football                 & LFR (0.3)        \\
Permanence    & {\bf (0.12,  -0.08)}    & {\bf (0.261,  0.091)}  & {\bf (0.121,  0.003)} \\
Closeness     & (0.019,  0.0048)  & (0.176,  0.041)   & (0.336,  0.301) \\
Betweenness     & (0.038,  0.043)  & (0.394,  0.358)   & (0.12,  0.1099) \\
Pagerank      & (0.0033,  0.0034) & (0.008,  0.008)   & (0.001,  0.001) \\
Modularity    & (0.467,  0.464)   & (0.555,  0.557)   & (0.661,  0.658) \\
Conductance & (0.552,  0.551)   & (0.407,  0.4003)   & (0.303,  0.3)   \\
CutRatio    & (0.012,  0.0098)  & (0.038,  0.027)  & (0.004,  0.003)\\
\end{tabular}
\\
\hline
\begin{tabular}{llll}
Permanence    & {\bf (0.116,  -0.028)}  & {\bf (0.355,  0.165)} & {\bf (0.109, -0.09)} \\
Closeness     & (0.02,  0.004)    & (0.176,  0.037)  & (0.336,  0.293) \\
Betweenness   & (0.041,  0.065)    & (0.39,  0.40)  & (0.162,  0.303) \\
Pagerank      & (0.0033,  0.0034) & (0.008,  0.008) & (0.001,  0.001) \\
Modularity    & (0.467,  0.502)   & (0.555,  0.548)  & (0.661,  0.659) \\
Conductance & (0.548,  0.513)   & (0.406,  0.401) & (0.303,  0.301) \\
CutRatio    & (0.012,  0.0098)  & (0.038,  0.027) & (0.004,  0.003)
\end{tabular}
\\
\hline
\begin{tabular}{llll}
Permanence    & {\bf (0.09,  -0.016) }  & {\bf(0.25,  0.066)} & {\bf (0.129, -0.097)} \\
Closeness     & (0.02,  0.001)    & (0.183,  0.047) & (0.336,  0.137)  \\
Betweenneness     & (0.038,  0.031)    & (0.391,  0.198) & (0.106,  0.072)  \\
Pagerank      & (0.0033,  0.0033) & (0.008,  0.008) & (0.001,  0.001)  \\
Modularity    & (0.451,  0.4)     & (0.549,  0.526) & (0.657,  0.644)  \\
Conductance & (0.583,  0.576)   & (0.407,  0.441) & (0.304,  0.327)  \\
CutRatio    & (0.012,  0.0098)  & (0.038,  0.027)  & (0.004,  0.003) 
\\
\hline
\end{tabular}
\end{tabular}}
\caption{Range of metric values for uniform noise ($1^\textrm{st}$ row), censored noise ($2^\textrm{nd}$ row) and crawled noise ($3^\textrm{rd}$ row). The numbers in bold shows the range that has the largest separation.}
\label{range}
\vspace{-.8cm}
\end{table}

\section{Reliability of the Metrics}\label{Section4}

In message spreading~\cite{chierichetti2010rumour}, a set of source vertices (seed nodes) start sending a message. At every time step, a vertex containing the message transfers the message uniformly at random to one of its neighbors who does not have the message. The algorithm terminates when
all vertices have received the message. The selection of the seed nodes is critical to how quickly the message spreads. A {\em reliable} metric is one whose high ranked nodes, if used as seeds, can spread the message quickly even under noise.

{\bf Methodology.} For each of the centrality metrics, closeness, betweeeness and Pagerank, and also for permanence we select a small fraction of the highest ranked nodes as the seed. We also select seeds (a) uniformly at random and (b) based on highest degree as baselines for the spreading experiments. 

 For different levels of noise, we compute the the number of iterations required to broadcast the message in the whole network and compare the values across the different centrality metrics, permanence and the two baselines.

\begin{figure}
\centering
  \begin{tabular}{l}
    \includegraphics[scale=.2]{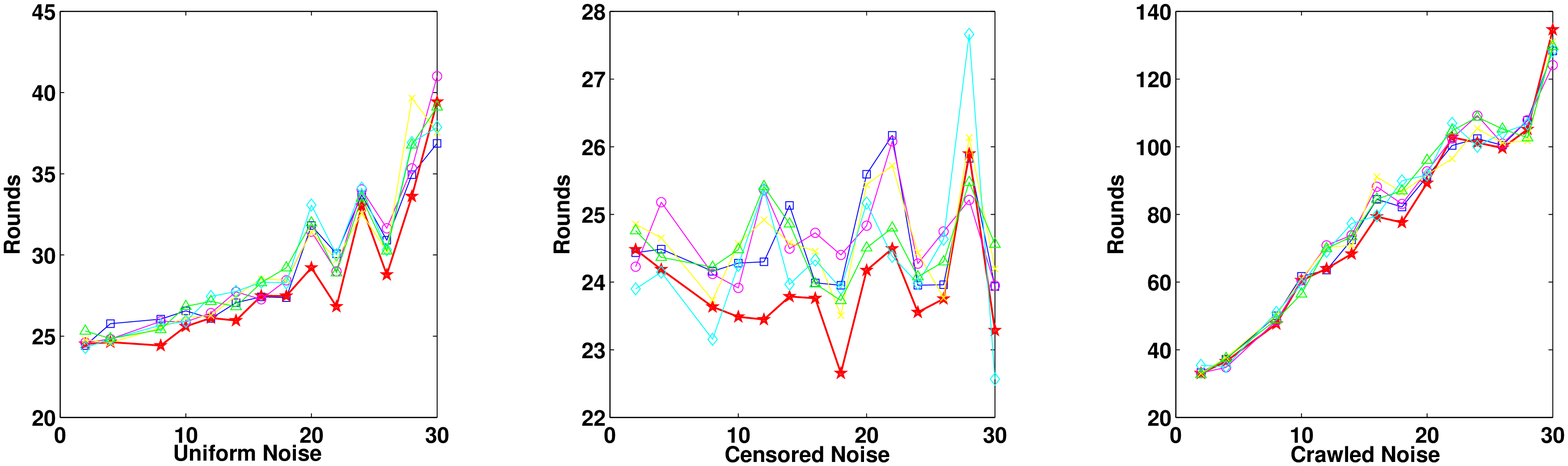} \\
    \includegraphics[scale=.2]{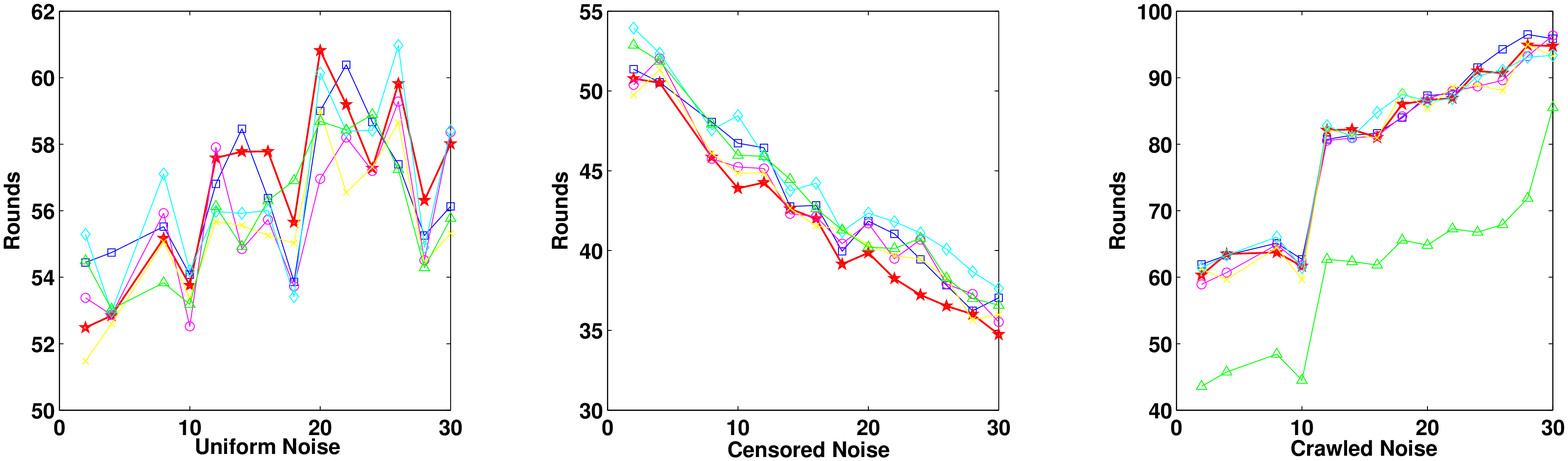} \\
 \includegraphics[scale=.2]{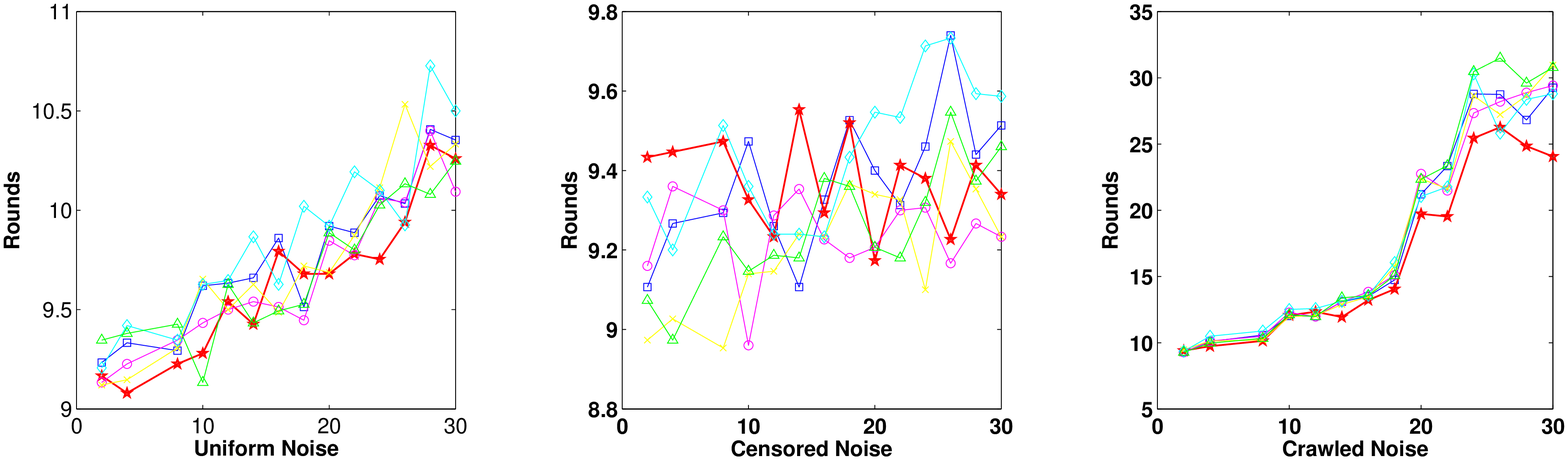} 
  \end{tabular}
  \caption{Time required to broadcast a message for different seed node selection mechanisms (permanence, closeness, betweenness, Pagerank) and varying noise levels (in steps of 2\%). LFR network ($\mu = 0.3$, first panel), railway network (second panel) and football network (third panel). The first column of figures represents results for uniform noise, the middle column represents results for censored noise and the last column represents results for the crawled noise.}
\label{spreading}
\vspace{-.5cm}
\end{figure}

{\bf Results:} In Fig.~\ref{spreading}, we plot the time required to broadcast for different levels of noise. For each noise level the results are averaged over ten different runs. The results can be divided into three groups.

{\em All metrics perform equally well:} Crawled noise in the LFR and the football network.

{\em Metrics perform differently based on the noise level, but no clear winner:} Uniform noise in the railway and football networks and censored noise in the football.

{\em One metric performs better in most of the noise levels:} For uniform and censored noise for the LFR network and censored noise for the railway network permanence takes the least time to spread messages. For crawled noise in the railway network, betweenness takes the least time.

We therefore see that for the . Therefore vertices with high betweenness centralities would be key connection points. This feature is exaggerated in the crawled noise since the network created using BFS-search has become tree-like.

\section{Analysis of performance}\label{Section5}
In this section, we explore the characteristics of permanence that make it such a strong measure under noise.

\begin{figure}
 \begin{tabular}{l}
    \includegraphics[scale=.17]{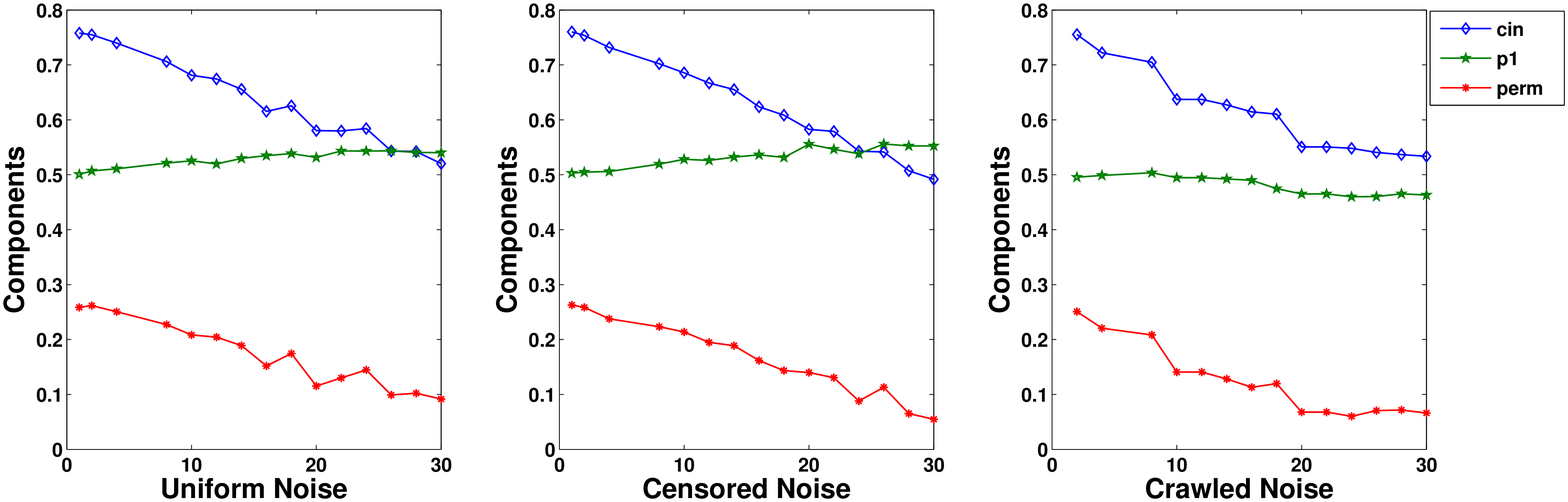} \\
    \includegraphics[scale=.17]{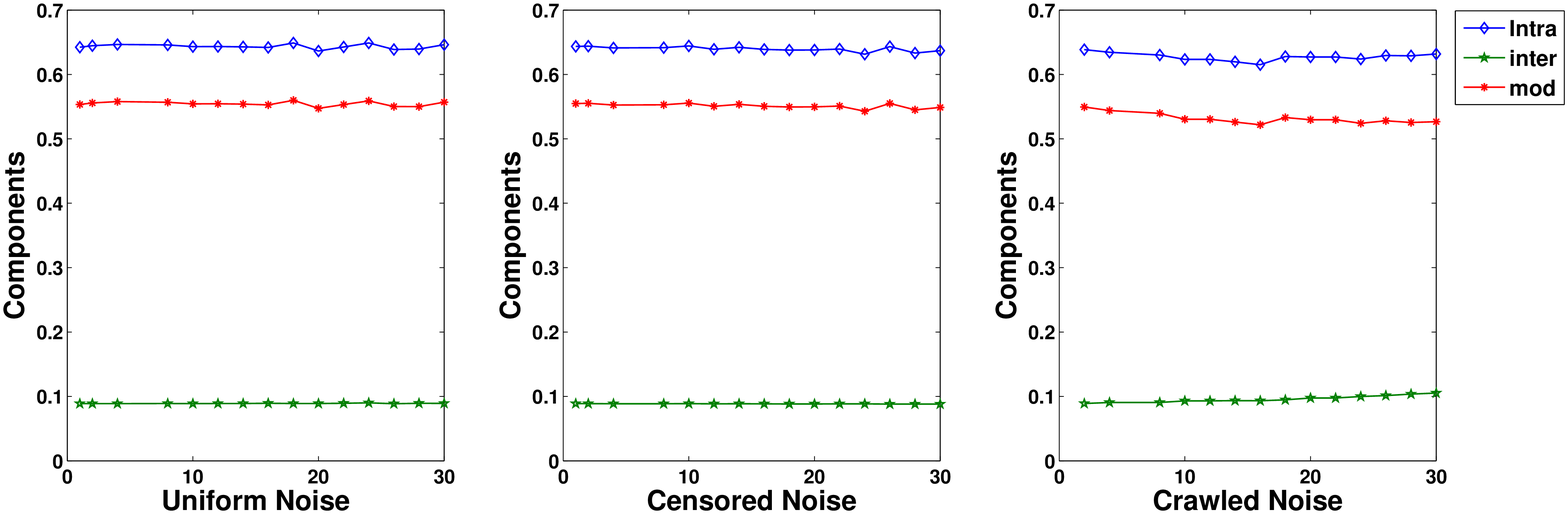} \\
  \end{tabular}
  \caption{The variation in the different components of permanence and modularity when the noise levels are varied for the football network.}
  \label{components1}
  \vspace{-.5cm}
\end{figure}

\subsection{Sensitivity of permanence}
We compare the sensitivity of permanence with other community scoring metrics.
We break the permanence formula into two parts $PI=\frac{I(v)}{E_{max}(v)} \times \frac{1}{D(v)}$ and $c_{in}(v)$, and observe how they change for the different noise models. The results in Fig.~\ref{components1} show that $PI$ remains relatively constant, whereas the internal clustering coefficient is the major contributor to the change in permanence. When we contrast this result with the main factors in modularity (Fig.~\ref{components1}), namely the internal and external edges, we see that each factor remains relatively constant. A similar observation holds when we consider the other scoring metrics conductance and cut ratio. 
 
 



\subsection {Rank of high permanence vertices under noise}
\begin{figure}[!ht]
\centering
  \begin{tabular}{c}
    \includegraphics[scale=.2]{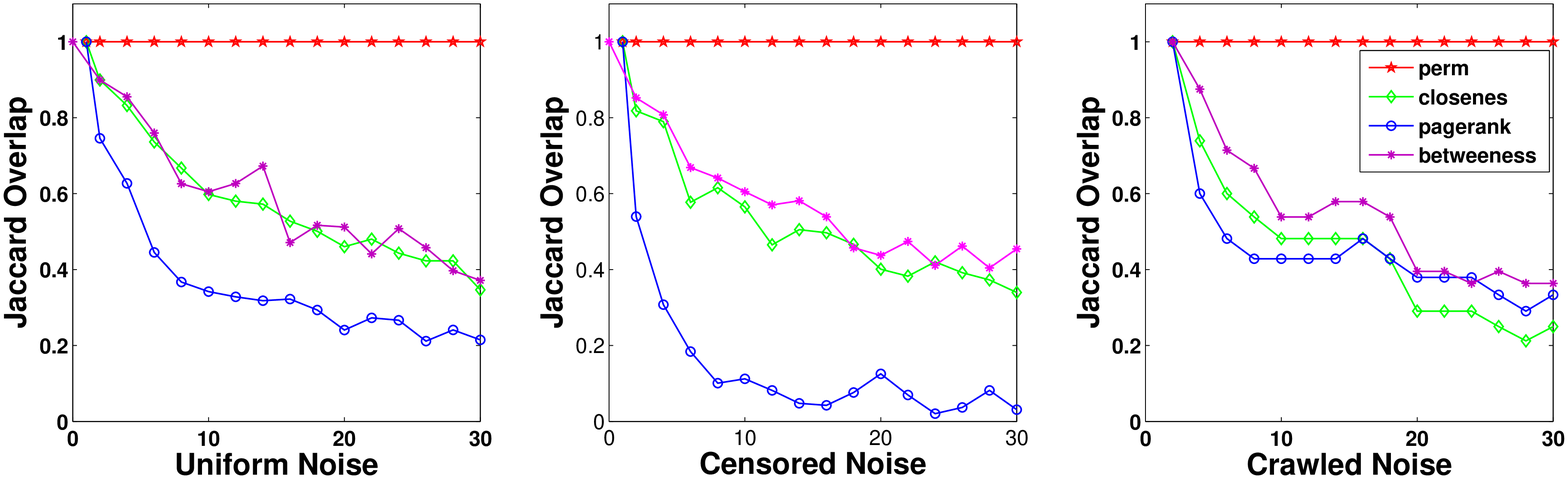}\\
    \includegraphics[scale=.2]{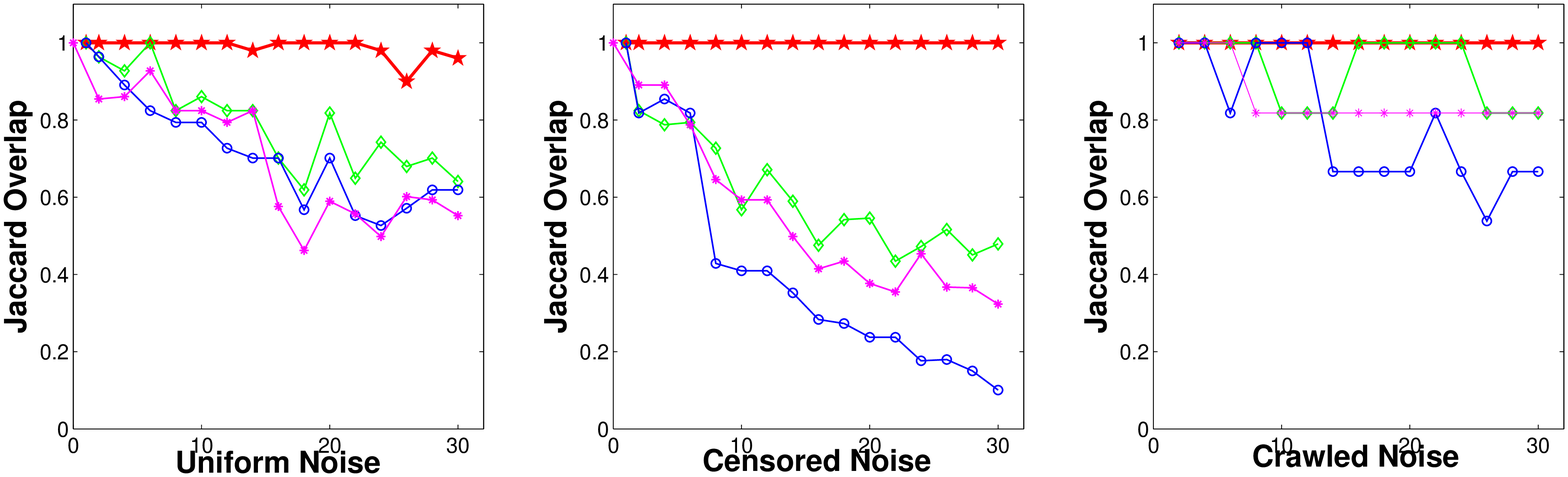} \\
    \includegraphics[scale=.2]{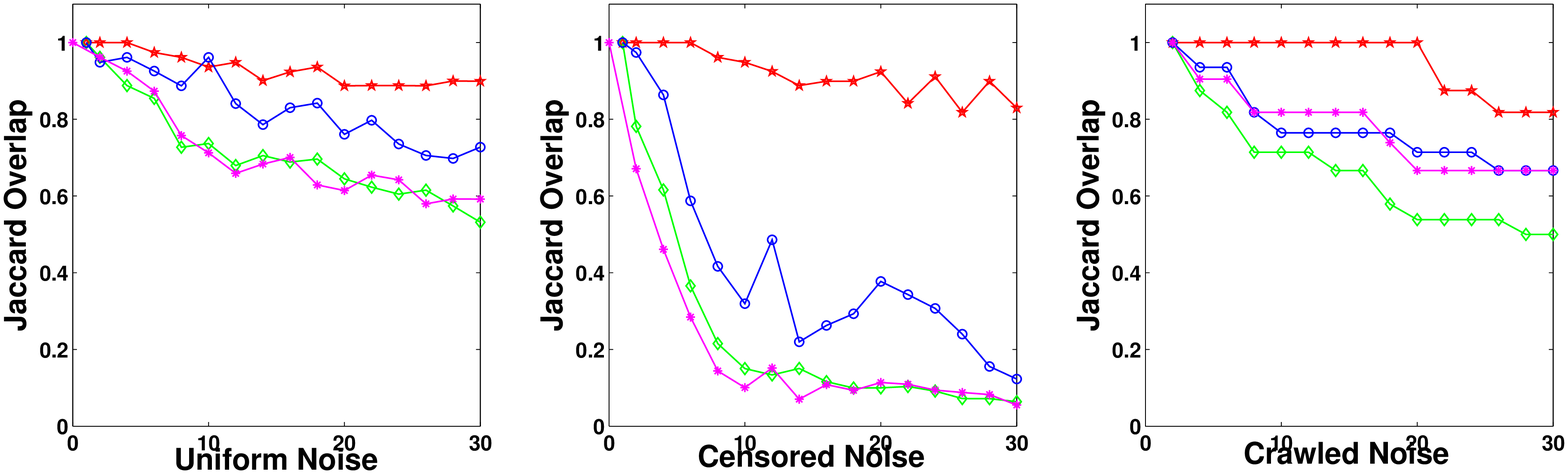}
  \end{tabular}
  \caption{The Jaccard Index between the top vertices of the original and the noisy networks for varying noise levels. We show the football (top), the railway (middle) and the LFR ($\mu=0.3$, bottom) networks respectively.}
  \label{rank}
  \vspace{-.5cm}
\end{figure}

We compare the centrality metrics and permanence to check how their top ranking vertices alter under noise models. We identify the top 20 of the high valued vertices for each  metric. Then for each noise level we compute the new top ranked vertices. We compute the Jaccard Index~\cite{gower1985measures} between the original vertex set and the new one obtained from the noisy network. A high Jaccard index (maximum value 1) indicates that most of the top ranked vertices are retained under noise, and a low value (minimum 0) indicates that the set has changed completely. As can be seen in Fig.~\ref{rank}, the Jaccard Index deteriorates much more slowly for permanence, than the other centrality metrics in most of the cases. This indicates that the ids of the high valued permanence vertices remain relatively constant under noise.


These experiments together provide a rationale of why permanence is effective both in evaluating noise through sensitivity and also reliable for message spreading.

\vspace{-.3cm}

\section{Conclusion}\label{Section7}
In this work, we have done rigorous experiments to understand the effect of noise in complex networks and compared different community scoring and centrality metrics in terms of sensitivity and reliability. A key observation is that in a majority of cases permanence worked better than all the other competing measures investigated. The central lesson is that while permanence is appropriately sensitive to different noise levels, the high permanence nodes are almost unaffected by the application of noise thus making the measure at the same time very reliable. 

In future we would like to investigate the analytical reasons for the stability of high permanence nodes and, thereby, propose an algorithm to automatically identify the level of noise up to which this stability persists. 

The data and the code are available in the public domain (\url{https://github.com/Sam131112/Noise_Models.git}).
\section* {Acknowledgment}
SS and AM acknowledges financial support from project DISARM, ITRA, DeiTY. SB acknowledges funding from NSF-CCF Award \# 153881.

%
\IEEEpeerreviewmaketitle

\bibliographystyle{abbrv}
\bibliography{main.bib}

\end{document}